\documentclass[12pt]{article}
\usepackage{epsfig}
\usepackage{dcolumn}
\usepackage{bm}
\usepackage{amssymb}
\usepackage{amsmath}
\usepackage{multirow}
\usepackage[english]{babel}
\usepackage[]{graphicx}
\usepackage{slashed}
\usepackage{hep,hepnicenames}
\newcommand{\bsg}{\ensuremath{\mathcal{B}(b \to s \gamma)}}
\newcommand{\CP}{\ensuremath{CP}}

\newcommand{\hzero}{\ensuremath{\PHiggslightzero}} 
\newcommand{\Hzero}{\ensuremath{\PHiggsheavyzero}} 
\newcommand{\Azero}{\ensuremath{\PHiggspszero}} 

\def\Title#1{\begin{center} {\Large {\bf #1} } \end{center}}
\usepackage{fancyhdr}
\begin{document}
\topskip 2cm

\Title{Single Higgs boson production at a photon-photon collider: a 2HDM/MSSM comparison}
\bigskip

\begin{raggedright}  

{\it \underline{David L\'opez-Val}  \index{}
\footnote{Presented at Linear Collider 2011: Understanding QCD at Linear Colliders 
in searching for old and new physics, 12-16 September 2011, ECT*, Trento, Italy}\\
Institut f\"ur Theoretische Physik,Universit\"at Heidelberg\\
Philosophenweg 16, D-69120 Heidelberg, Germany \\
{\rm  lopez@thphys.uni-heidelberg.de}
}\\
\vspace{0.2cm}

\bigskip\bigskip
\end{raggedright}
\vskip 0.5  cm
\begin{raggedright}
{\bf Abstract} 
We consider the loop-induced production of a single Higgs boson from direct
$\Pphoton\Pphoton$-scattering at a photon collider. A dedicated
analysis of the
total cross section $<\sigma_{\gamma\gamma \to h}>$ (for $h=\hzero,
\Hzero, \Azero$), and the relative strength of the effective $h\gamma\gamma$ coupling $r \equiv g_{\Pphoton\Pphoton h}/g_{\Pphoton\Pphoton H_{SM}}$,
is carried out
within the general Two-Higgs-Doublet Model (2HDM) and
the Minimal Supersymmetric Standard Model (MSSM). 
We systematically survey representative regions over the 
parameter space,
in full agreement with brought-to-date theoretical and phenomenological
restrictions, and obtain
production rates up
to $10^4$ Higgs boson events per $500 \invfb$ of integrated luminosity. We identify 
trademark phenomenological profiles for the different $\gamma\gamma \to h$ channels and trace them back
to the distinctive dynamical features characterizing each of these models -- most 
significantly, the enhancement potential of the Higgs self-interactions
in the general 2HDM. The upshot of our results illustrates the possibilities
of $\gamma\gamma$-physics and emphasizes the relevance of linear colliders
for the Higgs boson research program.
%
\end{raggedright}

\section{Introduction}
\label{sec:intro}

The LHC is now truly laying siege to the Higgs boson. The 
diphoton and gauge boson pair excesses recently reported by ATLAS
and CMS \cite{evidence} may indeed constitute, if confirmed, 
a first solid trace of its existence. In the meantime, the currently available data
keeps narrowing down
the mass range and the phenomenological portray under which the Higgs boson may manifest. On
the other hand, 
strong theoretical motivation supports of the idea that Electroweak Symmetry Breaking (EWSB)
is realized by some mechanism beyond that of the Standard Model (SM), 
viz. of a single, fundamental spinless field.
One canonical example of the latter is the general 2HDM \cite{2hdmrev}. Here, the addition of a second scalar
$SU_L(2)$ doublet tailors a rich and disclosing phenomenology \cite{recent2hdm}.
The 2HDM can be fully set along in terms of the the physical
Higgs boson masses; the ratio $\tan\beta \equiv \langle
H_2^0\rangle/\langle H_1^0\rangle$ of the two Vacuum Expectation Values (VEVs) giving masses to
the up- and down-like quarks; the mixing angle $\alpha$ between the
two $\CP$-even states, $\hzero,\Hzero$; and, finally, one genuine Higgs boson
self-coupling, which we shall denote $\lambda_5$. The Higgs sector of the MSSM
corresponds to a particular (supersymmetric) realization of the general (unconstrained) 2HDM \cite{mssm}.
For further details we refer the reader to 
Ref.\,\cite{loop1}, where all the notation, model setup
and restrictions are discussed at length.

Following the eventual discovery of the Higgs boson(s) at the LHC,
of crucial importance will be to address the precise experimental determination
of its quantum numbers, mass spectrum and couplings to other particles.
A linear collider (linac) can play a central role in this enterprise \cite{ILCPhysics}. 
Dedicated studies have exhaustively sought for the phenomenological
imprints of the basic 2HDM Higgs boson production modes,
such as e.g.
i) triple Higgs, $\APelectron\Pelectron \to 3h$ \cite{giancarlo};
ii) inclusive Higgs-pair through EW gauge boson fusion, $\APelectron\Pelectron \to V^*V^* \to 2h+X$ \cite{neil};
iii)  exclusive Higgs-pair $\APelectron\Pelectron  \to 2h$ \cite{loop1,hw};
and iv) associated Higgs/gauge boson $\APelectron\Pelectron  \to hV$ \cite{loop2}, 
with $h \equiv \hzero,\Azero,\Hzero,\PHiggs^{\pm}$ and $V  \equiv \PZ^0,\PW^{\pm}$
\footnote{For related work in the context of
MSSM Higgs boson production see e.g. ~\cite{mssmloop}.}. 
As a common highlight, all these studies report 
sizable production rates and large quantum effects,
arising from the potentially enhanced Higgs self-interactions. 
These self-couplings, unlike their MSSM analogues, are not anchored by the gauge
symmetry, and may thus be strengthened as much as allowed by the unitarity bounds. 
Interestingly enough, Higgs boson searches at an $\APelectron\Pelectron$ 
collider may benefit from alternative operation modes, particularly from $\Pphoton\Pphoton$
scattering. In this vein, single ($\gamma\gamma \to h$) and double ($\gamma\gamma \to 2h$) Higgs boson
pair production are examples of $\gamma\gamma$-induced processes which entirely operate at the quantum level.
The effective (loop-mediated) Higgs/photon interaction $g_{\Pphoton\Pphoton h}$ can be regarded
as a direct
probe of non-standard (charged) degrees of freedom coupled to the Higgs sector.
The aforementioned single Higgs channels have been considered in the framework of the SM \cite{photon_sm},
the 2HDM \cite{photon_2hdm} and the MSSM \cite{photon_mssm, previousmssm} and are known to exhibit excellent
experimental prospects, not only due to the clean environment inherent to a linac machine,
but also owing to the high attainable $\gamma\gamma$ luminosity, and the possibility to 
tune the $\Pphoton$-beam polarization as a strategy
to enlarge the signal-versus-background ratios\footnote{Analogue studies for the $\gamma\gamma\to hh$ mode
are available e.g. in Ref.~\cite{doublephoton}.}.

\section{Numerical analysis}
\label{sec:results}

\subsection{Computational setup}

\begin{figure}[t!]
\begin{center}
\begin{tabular}{ccc}
\includegraphics[scale=0.4]{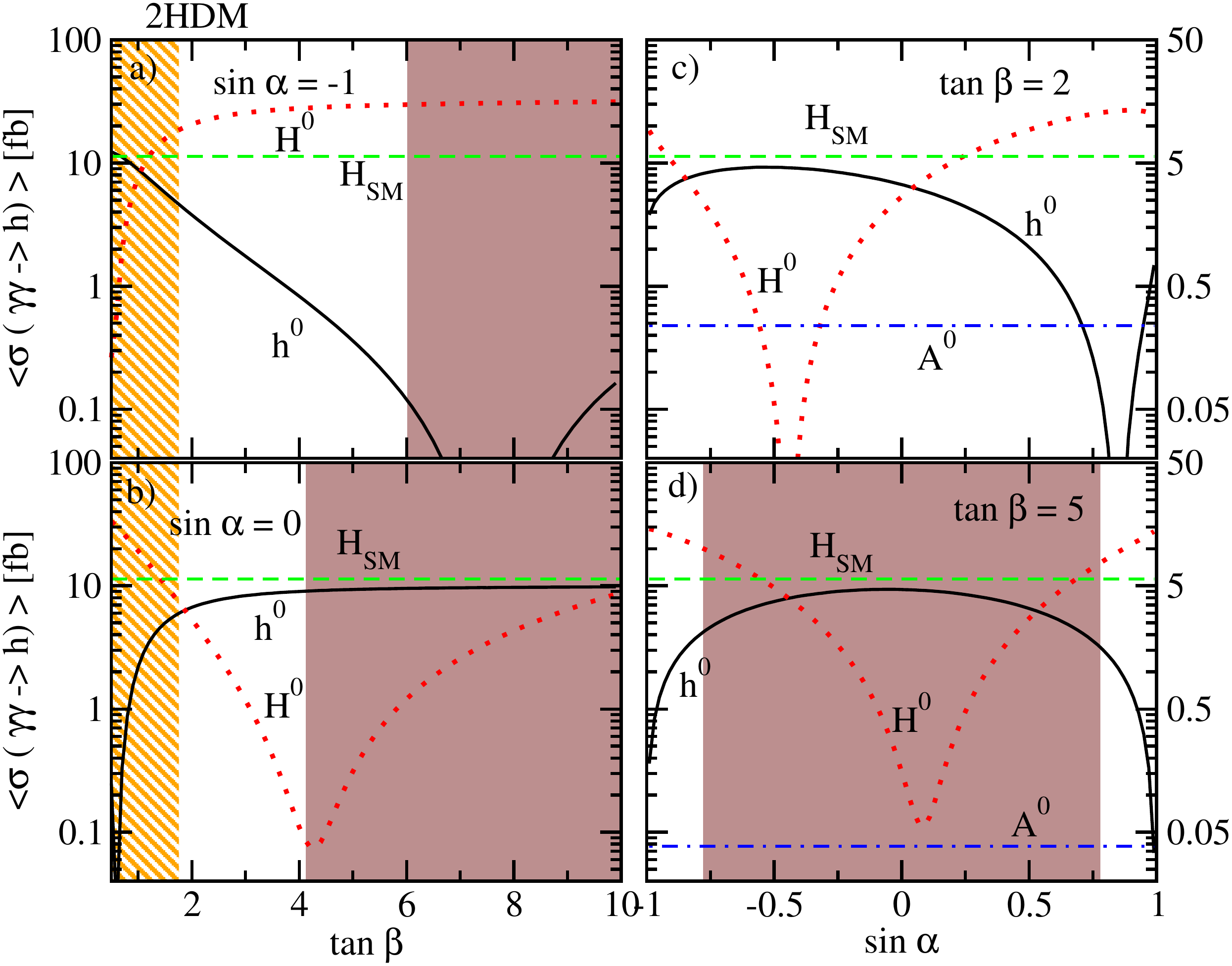} & & \includegraphics[scale=0.45]{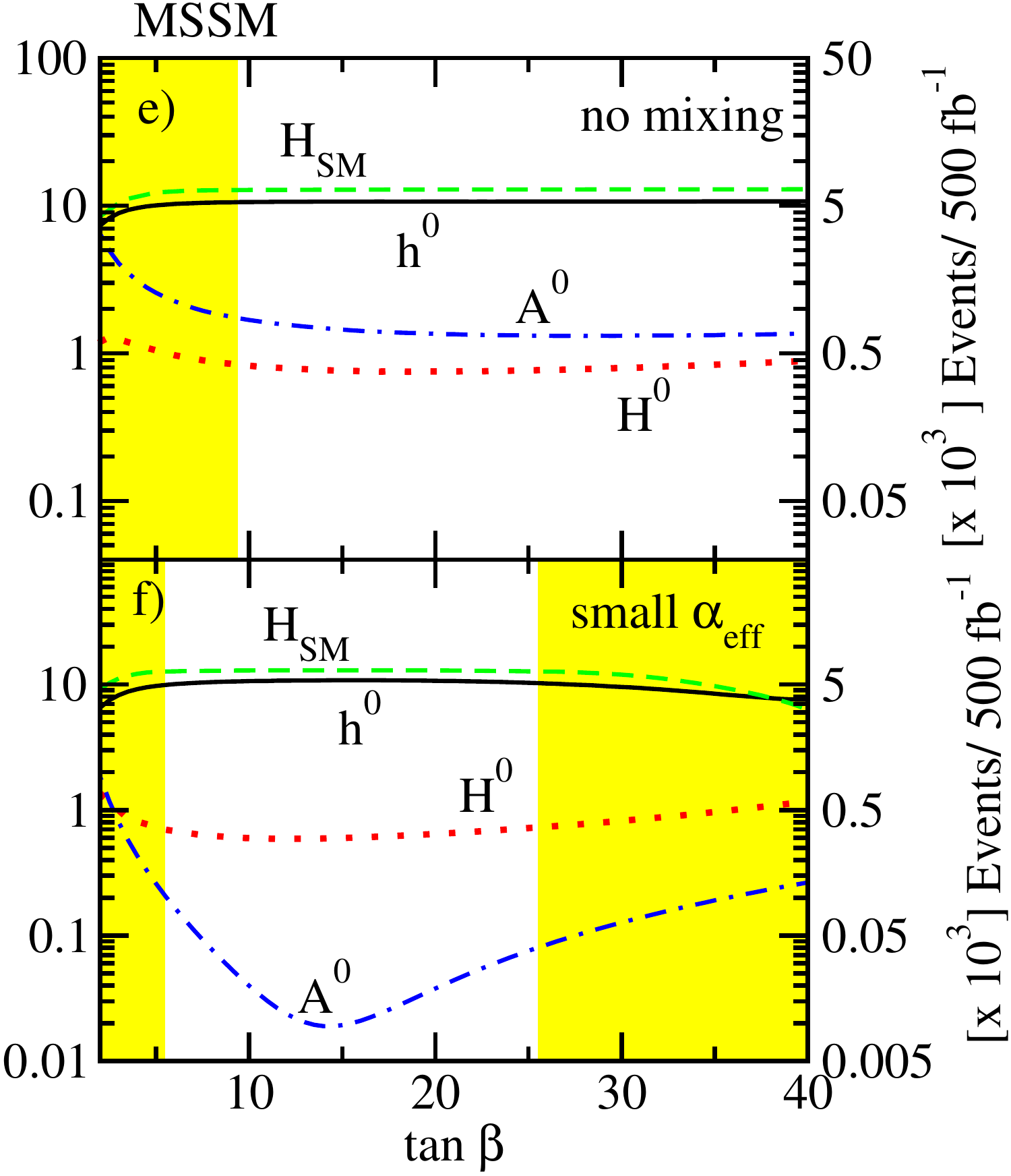}
\end{tabular}
\caption{\footnotesize{\textbf{Left panels (a-d):} Total spin-averaged cross-section
$\langle\sigma_{\gamma\gamma\to h}\rangle(s)$} 
\footnotesize{and number of Higgs boson events, as a function
of $\tan\beta$ (a,b) and $\sin\alpha$ (c,d) within the 2HDM. 
The shaded (resp. dashed) areas are excluded by unitarity
(resp. $B_d^0-\bar{B}_d^0$ mixing). The Higgs boson masses are fixed as follows:
$M_{\hzero} = 115 \,\GeV; M_{\Hzero} = 165 \,\GeV;M_{\Azero} = 100 \,\GeV;
M_{\PHiggs^{\pm}} = 105 \,\GeV$, with $\lambda_5 = 0$.
\textbf{Right panels (e-f):} 
$\langle\sigma_{\gamma\gamma\to h}\rangle(s)$ within the MSSM, as
a function of $\tan\beta$, for both the \emph{no-mixing} and the
\emph{small-$\alpha_{eff}$} benchmark points \cite{benchmarks}. The dashed regions
are ruled out by $b\to s \gamma$ data.
The linac center-of-mass energy is kept at $\sqrt{s} = 500 \,\GeV$.
}
}
\label{fig:survey}
\end{center}
\end{figure}

In this contribution we present a fully updated analysis of the process $\gamma\gamma\to h \; (h=\hzero,
\Hzero, \Azero)$ and undertake a comparison of the 2HDM -- versus the MSSM results. 
We focus our attention on the following 
two quantities: \textbf{i)}
the total, spin-averaged cross section,

\vspace{-0.5cm}
\begin{eqnarray}
\langle\sigma_{\gamma\gamma\to h}\rangle(s) &=& \sum_{\{ij\}}
\int_{0}^1\,d\tau\,\frac{d\,\mathcal{L}_{ij}^{ee}}{d\tau}\,\hat{\sigma}_{\eta_i\,\eta_j}(\hat{s})\,,
\label{eq:sigmatotal}
\end{eqnarray}
\vspace{-1.5cm}

\noindent where $\hat{\sigma}_{\eta_i\,\eta_j}$ stands for the ``hard'' scattering cross section,
$\hat{s}=\tau\,s$ being the partonic center-of-mass energy; while  
$d\,\mathcal{L}_{ij}^{ee}/d\tau$ denotes the (differential)
photon luminosity distributions, by which we describe the effective $e^{\pm} \to \gamma$ conversion
of the primary linac beam. In turn, $\eta_{i,j}$ accounts for the respective polarization
of the resulting photon beams; and \textbf{ii)}
the $\Pphoton\Pphoton h$ coupling
strength, $r \equiv g_{\Pphoton\Pphoton h}/g_{\Pphoton\Pphoton H_{SM}}$ -- that we normalize
to the SM, identifying $\hzero \equiv \PHiggs_{SM}$.
We compare the distinct phenomenological
patterns that emerge from the 2HDM and the MSSM
and spell out the specific dynamical features that may help to disentangle
both models. Further details may be found in Refs. \cite{photon_2hdm,photon_mssm}.  

Throughout our study we make use of the standard algebraic and numerical
packages {\sc FeynArts, FormCalc} and {\sc LoopTools} \cite{hahn}. 
Updated
experimental constraints (
stemming from the EW precision data, low-energy flavor-physics 
and the Higgs mass regions ruled out by the LEP, Tevatron and LHC direct
searches), 
as well as the theoretical consistency
conditions (to wit: 
perturbativity, unitarity and vacuum stability) are duly taken into account -- cf.
~\cite{constraints_general,superiso,unitarity,vacuum,2hdmcalc,higgsbounds}. The photon luminosity distributions
are obtained from \cite{compaz}, while the MSSM Higgs mass spectrum is provided by 
{\sc FeynHiggs}\,\cite{feynhiggs}.

\subsection{Profiling $\gamma\gamma \to h$ within the 2HDM}

\begin{figure}[t!]
\begin{center}
\begin{tabular}{ccc}
\includegraphics[scale=0.75]{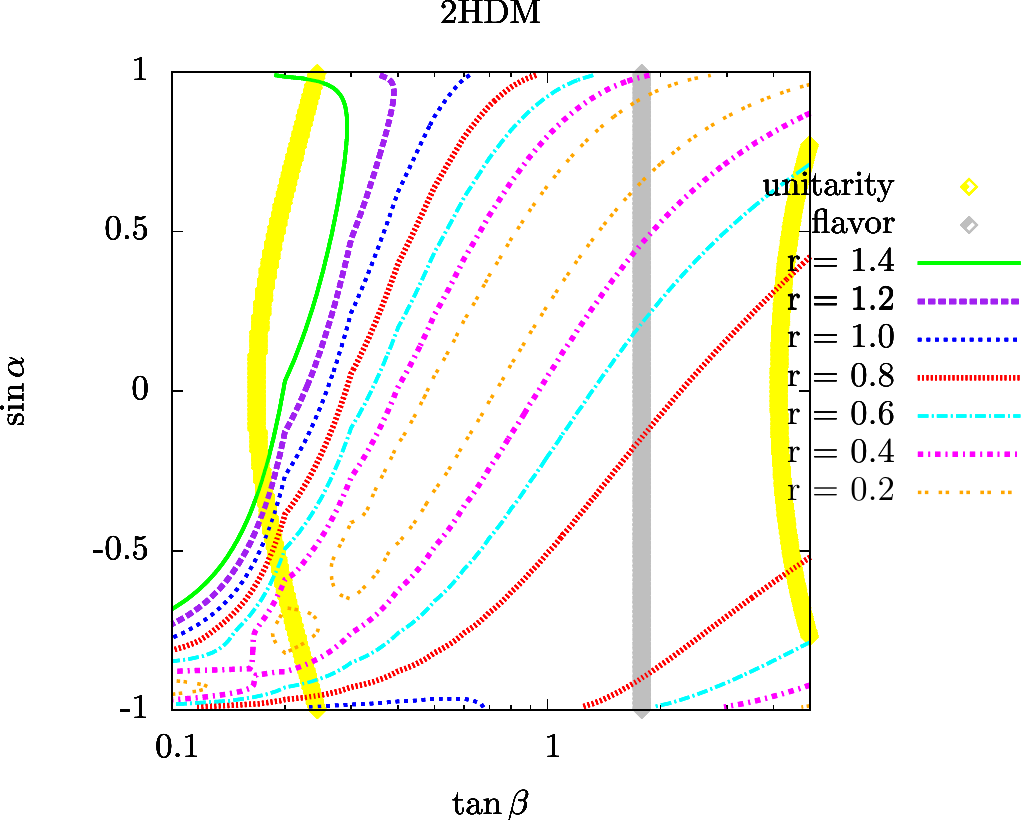} & & \includegraphics[scale=0.7]{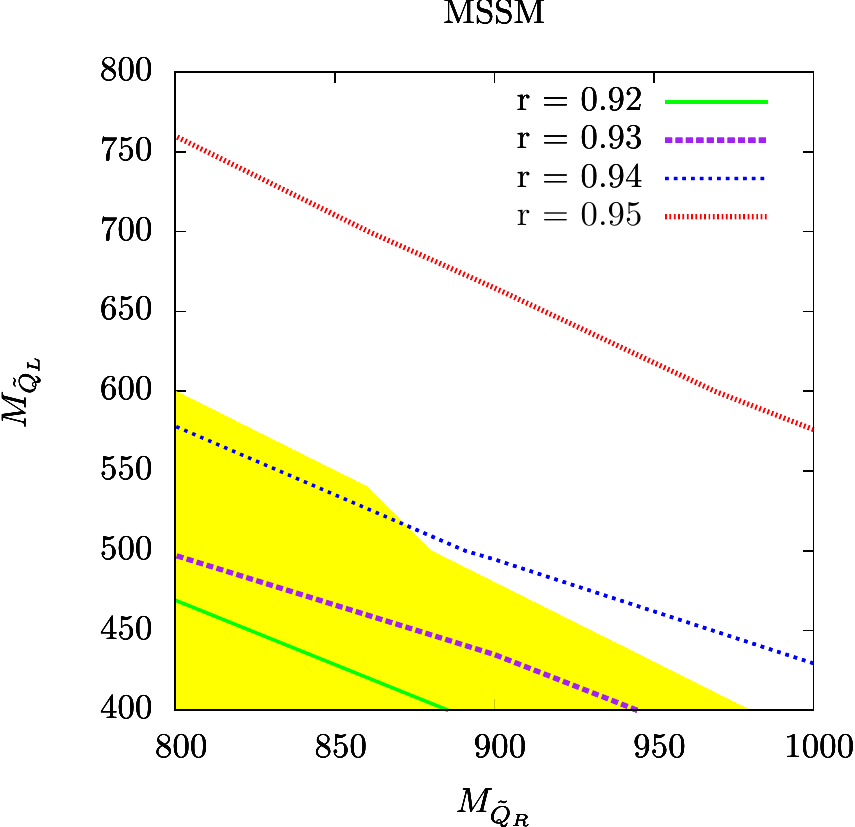}
\end{tabular}
\caption{\footnotesize{Contour plots of the
ratio $r \equiv g_{\Pphoton\Pphoton \hzero}/g_{\Pphoton\Pphoton H_{SM}}$
that measures the effective $\Pphoton\Pphoton \hzero$ coupling strength normalized
to the SM, for representative parameter space configurations, comparing the
2HDM (left panel) and MSSM (right panel). The 2HDM calculation
is carried out assuming type-I Higgs/fermion Yukawa couplings, $\lambda_5 = 0$ and the same
set of Higgs boson masses as in Fig.~\ref{fig:survey}. 
The yellow strips on the left plot denote the lower and upper bounds ensuing from unitarity, while
the grey vertical band displays the restrictions stemming from $B_d^0-\bar{B}_d^0$. As for
the MSSM parameter setup, we employ
$\tan\beta = 2$, $M_{\Azero} = 600 $ GeV, $\mu = 500 $ GeV, $A_t =
1800$ GeV, $M_2 = 500$ GeV. The dashed area
is ruled out by $b\to s\gamma$. The linac center-of-mass energy is kept at $\sqrt{s} = 500 \,\GeV$.
}
}
\label{fig:scan}
\end{center}
\end{figure}

The upshot of our numerical analysis is displayed on the left
panels of Figs.~\ref{fig:survey} - \ref{fig:scan}.  There we illustrate
the behavior of $<\sigma_{\gamma\gamma\to h}>$ and the ratio $r$ over representative regions
of the 2HDM parameter space. For definiteness, we perform our calculation for a
type-I 2HDM structure and for relatively light Higgs boson masses (as quoted in the Figure
caption). The pinpointed trends, however, do not critically depend 
on the previous assumptions -- see Ref.~\cite{photon_2hdm,photon_mssm} for an extended discussion.
Our results neatly illustrate the interplay of the charged Higgs boson, $\PW^\pm$ boson
and fermion loops, whose respective contributions to $g_{\Pphoton\Pphoton h}$
undergo a highly characteristic destructive interference. 
The strength of the Higgs self-coupling $\lambda_{h\PHiggs^+\PHiggs^-}$, which is primarily
modulated by $\tan\beta$ and $\lambda_5$, determines whether the overall rates may become enhanced
($r > 1$) or suppressed ($r<1$) relatively to the SM expectations. Scenarios yielding $r>1$ could in principle
be met for $\lambda_{h\PHiggs^+\PHiggs^-} \sim \mathcal{O}(10^3)$ GeV and $M_{\PHiggs^{\pm}} \sim \mathcal{O}(100) \, \GeV$
(due to strongly boosted $\PHiggs^{\pm}$-mediated
loops) or $\tan\beta < 1$ (which enhances the top-mediated loops through the 
Higgs-top Yukawa coupling, $g_{\hzero t\bar{t}} \sim 
\sin\alpha/\sin\beta$). 
In practice, however, both situations are
disfavored by the combined effect of the unitarity and vacuum stability conditions, together 
with the flavor physics constraints (mostly from $B_d^0 - \bar{B}_d^0$). 
Instead, the 2HDM regions with $\lambda_{h\PHiggs^+\PHiggs^-} \sim \mathcal{O}(10^2) \,\GeV$ 
give rise to a trademark suppression of the single Higgs
boson rates, and pull the relative $h\Pphoton\Pphoton$ coupling strength down
to values of $r \sim -50\%$. Away from these
largely subdued domains, we retrieve
total cross sections in the ballpark of $<\sigma_{\gamma\gamma \to h}> \sim 1-50$ fb -- this is to say,
up to a few thousand neutral, \CP-even, single Higgs
boson events, for the light ($\hzero$) and the heavy ($\Hzero$) states alike.
Finally, if the Higgs
self-interactions are even weaker -- or, alternatively, the charged Higgs bosons are very massive --
then the $\PHiggs^{\pm}$-mediated corrections become subleading. In such instances we are left with $r \lesssim 1$,
as a reflect of the fact that the $g_{\Pphoton\Pphoton h}$ coupling is now essentially determined by a SM-like combination
of $\PW^{\pm}$ and fermion-mediated loops. It is also worth underlining the complementary nature of the 
production rates for the two neutral \CP-even Higgs channels $\gamma\gamma \to \hzero/\Hzero$, which
ensues from the inverse correlation of the respective couplings to the charged Higgs, 
namely of $\lambda_{ \hzero\PHiggs^+\PHiggs^-}$ with respect to  
$\lambda_{ \Hzero\PHiggs^+\PHiggs^-}$ -- see the $\sigma_{\hzero}$ and $\sigma_{\Hzero}$ curves
from panels a-d in Fig.~\ref{fig:survey}. 
We also observe
that the results for $\gamma\gamma \to \Hzero$ tend to be slightly above the SM yields, 
whereas $\gamma\gamma \to \hzero$ stays usually below. This follows from the kinematic
structure of the total cross section, 
$<\sigma_{\gamma\gamma \to h}> \sim M^4_{M_{h}}/M^2_W$, which implies $\sigma_{\Hzero} > \sigma_{\hzero}$
as $M_{\Hzero} > M_{\hzero} \equiv M_{\PHiggs_{SM}}$.
In contrast, and owing
to its \CP-odd nature, $\gamma\gamma \to \Azero$ is essentially featureless and entails a minor
numerical impact. 

\subsection{Profiling $\gamma\gamma \to h$ within the MSSM}

Let us now turn our attention to the MSSM.
On the right panels of Figs.~\ref{fig:survey}-\ref{fig:scan} we survey the
behavior of the purported quantities $<\sigma_{\gamma\gamma \to h}>$ and $r$ 
for the representative MSSM parameter
setups that are quoted below \cite{benchmarks}:

\smallskip
{ \footnotesize
\begin{tabular}{ccccccc}
 & $M_{\Azero}\,[\GeV]$  &  $M_{SUSY}\,[\GeV]$ & $\mu\, [\GeV]$ & $X_t \equiv A_t-\mu/\tan\beta\, [\GeV]$& $M_2\,[\GeV]$ & $M_3\,[\GeV]$ \\
no-mixing & 400 & 2000 & 200 & 0 & 200 & 1600 \\
small $\alpha_eff$ & 300 & 800 & 2000 & -1100 & 500 & 500
\end{tabular}
}

\smallskip

\noindent We note that GUT relations {between $M_1$ and $M_2$},
as well as universal trilinear couplings ($A_t=A_b=A_\tau$), are
assumed throughout. Likewise, we duly account for the impact of the different sets of constraints, most significantly
stemming from $\bsg$ (dashed areas, in yellow) and the Higgs boson and squark mass bounds settled
by direct exclusion limits.

In this SUSY setup, non-standard contributions to the effective $g_{h\gamma\gamma}$ interaction
may emerge from a twofold origin. On the one hand we have a panoply of the 2HDM
one-loop diagrams mediated by the interchange of virtual charged Higgs bosons. In the present
framework, however, these terms do no longer bear any enhancement capabilities,
since the corresponding Higgs self-interactions are completely tied to the gauge couplings --
as a consequence of the underlying SUSY invariance. On the other
hand we find the squark-mediated quantum corrections. Their imprints on $g_{\Pphoton\Pphoton h}$ 
are mostly visible for relatively light squarks (with masses of few hundred GeV),
hand in hand with sizable mass splittings between their respective left and right-handed components
and large trilinear couplings to the Higgs bosons \footnote{The phenomenological implications
of this kind of Yukawa, and Yukawa-like couplings have been addressed in the past
in a wide variety of processes, see e.g. \cite{othersmssm}.}. In practice, however, the combination
of the different experimental restrictions effectively tames the abovementioned enlargement power.

We can thus conclude that realistic MSSM scenarios
encompass rather mild departures from the SM loop-induced mechanism
($r \sim -5 \% $), rendering overall production rates again in the ballpark of 
$<\sigma_{\gamma\gamma \to h}> \sim \mathcal{O}(10) \,$ fb for the lightest \CP-even state $h = \hzero$-- 
while its heavier companions $\Hzero,\Azero$ lie typically one order of magnitude below ~\cite{photon_mssm}.

\section{Discussion and concluding remarks}
\label{sec:discussion}

In this contribution we have reported on the single Higgs boson production through $\gamma\gamma$ scattering
in a TeV-range linear collider. The process $\gamma\gamma \to h$ is driven by an effective, loop-induced
$h\Pphoton\Pphoton$ interaction, a mechanism that is directly sensitive
to the eventual presence of new charged degrees of freedom. We have computed the total
cross section, $<\sigma_{\gamma\gamma \to h}>$, alongside with the effective (normalized) coupling
strength $r\equiv g_{\Pphoton\Pphoton h}/g_{\Pphoton\Pphoton H_{SM}}$, 
within both the 2HDM and the MSSM. We have disclosed characteristic phenomenological profiles and spelt out their
main differences, which mostly
stem from the respective Higgs self-interaction structures. In the MSSM, the aforementioned self-couplings
are anchored by the gauge symmetry, while in the 2HDM they can be as large as permitted
by the combined set of experimental and theoretical restrictions -- most significantly unitarity.
We have identified a sizable depletion of $<\sigma_{\gamma\gamma \to h}>$ (corresponding to values
of $r \sim -50\%$) for those 2HDM configurations in which a relatively
large $\lambda_{h\PHiggs^+\PHiggs^-}$ interaction is capable to thrust the $\PHiggs^{\pm}$-mediated contribution to $g_{\gamma\gamma h}$,
and subsequently to maximize the destructive interference that operates
between the different $\PHiggs^{\pm}$, $\PW^+$ and fermion-mediated loops. A smoking gun of underlying 2HDM physics
would thus manifest here as a missing number of single Higgs boson events.
On the MSSM side, 
departures from the SM are comparably much tempered ($ r \simeq -5 \%$) and essentially driven
by the squark-mediated corrections, which are relatively suppressed by the mass scale
of the exchanged SUSY particles and further weakened by the stringent
experimental bounds. An additional distinctive feature of both models might manifest
from the simultaneous observation of $\gamma\gamma \to \hzero$ and $\gamma\gamma \to \Hzero$.
Situations where both channels yield $\mathcal{O}(10^3)$ events per 500 $\invfb$ could only be attributed
to a non-standard, non-SUSY Higgs sector, since the mass splitting between the two
neutral, \CP-even Higgs states is typically enforced to be larger in the MSSM -- so that the
corresponding $\Pphoton\Pphoton \to \Hzero$ rates are comparably smaller.

The clean environment of a linac offers excellent prospects for the tagging and identification of
the single Higgs boson final states through the corresponding decay products. The latter
should arise in the form of either i) highly energetic, back-to-back heavy-quark dijets
($ h  \to jj$, with $jj \equiv c\bar{c},b\bar{b}$); ii) lepton tracks from gauge boson decays
($ h  \to \PW^+\PW^- \to 2l + \slashed{E}_T, \PZ^0\PZ^0 \to 4l$); iii) in the specific case of the 
MSSM, and if kinematically allowed, also the Higgs decays into
chargino pairs ($h \to \tilde{\chi_1}\tilde{\chi_2} \to jj + \slashed{E}_T $).
Precise Higgs boson mass measurements could then be conducted upon the reconstruction of the dijet -- or dilepton --
invariant masses and should broaden the present coverage of the LHC. For instance, they would enable to
sidestep the so-called ``LHC wedge'', 
namely the $M_{\Azero} \gtrsim 200 \GeV$ and $\tan\beta \sim \mathcal{O}(10)$ domains of the 
MSSM parameter space \cite{haberrev}. The dominant
backgrounds, corresponding to the processes $\gamma\gamma \to b\bar{b}/\PW^+\PW^-$, could be handled not only
by means of standard kinematic cuts, but also through a suitable tuning of the photon beam
polarization \cite{previousmssm}.

A future generation of linac machines, and of $\gamma\gamma$ facilities in particular,
should therefore be instrumental for a precise experimental reconstruction of the EWSB mechanism; namely
for the measurement of the Higgs boson mass, couplings and quantum numbers, if not for the discovery of
the Higgs boson itself
-- if its mass and/or its coupling pattern fell beyond the reach of the LHC and the $\APelectron\Pelectron$ colliders. 
Photon-photon physics may well furnish a most fruitful arena in which to carry the Higgs boson research program to completion. 

\bigskip
%
\textbf{Acknowledgements} It is a pleasure to thank Joan Sol\`a for 
the fruitful and enduring collaboration over the past years. I would also
like to express my gratitude to the organizers of the LC
2011 workshop at ETC-Trento for the
kind invitation to present this review, and for the kind atmospheare and enlightening
time we all shared at the meeting. 

\newcommand{\JHEP}[3]{ {JHEP} {#1} (#2)  {#3}}
\newcommand{\NPB}[3]{{\sl Nucl. Phys. } {\bf B#1} (#2)  {#3}}
\newcommand{\NPPS}[3]{{\sl Nucl. Phys. Proc. Supp. } {\bf #1} (#2)  {#3}}
\newcommand{\PRD}[3]{{\sl Phys. Rev. } {\bf D#1} (#2)   {#3}}
\newcommand{\PLB}[3]{{\sl Phys. Lett. } {\bf B#1} (#2)  {#3}}
\newcommand{\EPJ}[3]{{\sl Eur. Phys. J } {\bf C#1} (#2)  {#3}}
\newcommand{\PR}[3]{{\sl Phys. Rept. } {\bf #1} (#2)  {#3}}
\newcommand{\RMP}[3]{{\sl Rev. Mod. Phys. } {\bf #1} (#2)  {#3}}
\newcommand{\IJMP}[3]{{\sl Int. J. of Mod. Phys. } {\bf #1} (#2)  {#3}}
\newcommand{\PRL}[3]{{\sl Phys. Rev. Lett. } {\bf #1} (#2) {#3}}
\newcommand{\ZFP}[3]{{\sl Zeitsch. f. Physik } {\bf C#1} (#2)  {#3}}
\newcommand{\MPLA}[3]{{\sl Mod. Phys. Lett. } {\bf A#1} (#2) {#3}}
\newcommand{\JPG}[3]{{\sl J. Phys.} {\bf G#1} (#2)  {#3}}
\newcommand{\JPCF}[3]{{\sl J. Phys. Conf. Ser.} {\bf G#1} (#2)  {#3}}
\newcommand{\FDP}[3]{{\sl Fortsch. Phys.} {\bf G#1} (#2)  {#3}}
\newcommand{\CPC}[3]{{\sl Com. Phys. Comm.} {\bf G#1} (#2)  {#3}}

 \end{document}